%% file: Bi4Br4_ORNL-Draft_arXiv.tex
\documentclass[%
reprint,
superscriptaddress,
amsmath,amssymb,
aps,
prmaterials,
]{revtex4-2}

\usepackage[utf8]{inputenc}
\usepackage[T1]{fontenc}
\usepackage[english]{babel}

\usepackage{dcolumn}
\usepackage{bm}

\usepackage{amsfonts}

\usepackage{graphicx}

\usepackage{siunitx}

\usepackage{chemformula}

\usepackage{csquotes}

\usepackage{hyperref}

\hypersetup{
	breaklinks=true,   
	colorlinks=true,   
	pdfusetitle=true,  
}

\graphicspath{{./IMG/}}

\usepackage{xspace}

\newcommand{\aBiBr}{$\alpha$-\ch{Bi4Br4}\xspace}
\newcommand{\BiBr}{\ch{Bi4Br4}\xspace}

\begin{document}
	\title{Shear-resistant topology in quasi one-dimensional van der Waals material \ch{Bi4Br4}
 }
	
	\author{Jonathan K. Hofmann}
    \email{jo.hofmann@fz-juelich.de}
	\affiliation{Peter Grünberg Institut (PGI-3), Forschungszentrum Jülich, 52425 Jülich, Germany}
    \affiliation{Jülich Aachen Research Alliance (JARA), Fundamental of Future Information Technology, 52425 Jülich, Germany}
	\affiliation{Lehrstuhl für Experimentalphysik IV A, RWTH Aachen University, Otto-Blumenthal-Straße, 52074 Aachen,Germany}

	\author{Hoyeon Jeon}
	\author{Saban M. Hus}
	
	\affiliation{Center for Nanophase Materials Science, Oak Ridge National Laboratory, 
		Tennessee, USA}
    \author{Yuqi Zhang} 
    \affiliation{Key Laboratory of Advanced Optoelectronic Quantum Architecture and Measurement, Ministry of Education, School of Physics, Beijing Institute of Technology, Beijing 100081, China} 
    \affiliation{Beijing Key Lab of Nanophotonics and Ultrafine Optoelectronic Systems, Beijing Institute of Technology, Beijing 100081, China} 
    \affiliation{International Center for Quantum Materials, Beijing Institute of Technology, Zhuhai, 319000, China}

	\author{Mingqian Zheng}
	\affiliation{Centre for Quantum Physics, Key Laboratory of Advanced Optoelectronic Quantum Architecture and Measurement (MOE), School of Physics, Beijing Institute of Technology, Beijing 100081, China}

    \author{ Tobias Wichmann}
\affiliation{Peter Grünberg Institut (PGI-3), Forschungszentrum Jülich, 52425 Jülich, Germany}
\affiliation{Jülich Aachen Research Alliance (JARA), Fundamental of Future Information Technology, 52425 Jülich, Germany}
\affiliation{Lehrstuhl für Experimentalphysik IV A, RWTH Aachen University, Otto-Blumenthal-Straße, 52074 Aachen,Germany}

\author{An-Ping Li}
	
	\affiliation{Center for Nanophase Materials Science, Oak Ridge National Laboratory, 
		Tennessee, USA}

	\author{Jin-Jian Zhou}
	\affiliation{Centre for Quantum Physics, Key Laboratory of Advanced Optoelectronic Quantum Architecture and Measurement (MOE), School of Physics, Beijing Institute of Technology, Beijing 100081, China}
	
	\author{Zhiwei Wang}
	\author{Yugui Yao}
	\affiliation{Key Laboratory of Advanced Optoelectronic Quantum Architecture and Measurement, Ministry of Education, School of Physics, Beijing Institute of Technology, Beijing 100081, China} 
    \affiliation{Beijing Key Lab of Nanophotonics and Ultrafine Optoelectronic Systems, Beijing Institute of Technology, Beijing 100081, China}
    \affiliation{International Center for Quantum Materials, Beijing Institute of Technology, Zhuhai, 319000, China}

	\author{Bert Voigtländer}
	\author{F. Stefan Tautz}
	\affiliation{Peter Grünberg Institut (PGI-3), Forschungszentrum Jülich, 52425 Jülich, Germany}
    \affiliation{Jülich Aachen Research Alliance (JARA), Fundamental of Future Information Technology, 52425 Jülich, Germany}
	\affiliation{Lehrstuhl für Experimentalphysik IV A, RWTH Aachen University, Otto-Blumenthal-Straße, 52074 Aachen,Germany}
	
	\author{Felix Lüpke}
    \email{f.luepke@fz-juelich.de}
    
	\affiliation{Peter Grünberg Institut (PGI-3), Forschungszentrum Jülich, 52425 Jülich, Germany}
    \affiliation{II. Physikalisches Institut, Universit\"at zu K\"oln, Z\"ulpicher Straße 77, 50937 Köln, Germany}



\date{\today}
\begin{abstract}

Bi$_4$Br$_4$ is a prototypical quasi one-dimensional (1D) material in which covalently bonded bismuth bromide chains are arranged in parallel, side-by-side and layer-by-layer, with van der Waals (vdW) gaps in between.
So far, two different structures have been reported for this compound, \aBiBr and $\beta$-\BiBr, in both of which neighboring chains are shifted by $\mathbf{b}/2$, i.e., half a unit cell vector in the plane, but which differ in their vertical stacking.
While the different layer arrangements are known to result in distinct electronic properties, the effect of possible in-plane shifts between the atomic chains remains an open question.
Here, using scanning tunneling microscopy and spectroscopy (STM/STS), we report a new \ch{Bi4Br4}(001) structure, with a shift of $\mathbf{b}/3$ between neighboring chains in the plane and AB layer stacking.
We determine shear strain to be the origin of this new structure, which can readily result in shifts of neighboring atomic chains because of the weak inter-chain bonding. 
For the observed $b/3$ structure, the (residual) atomic chain shift corresponds to an in-plane shear strain of $\gamma\approx7.5\%$.
STS reveals a bulk insulating gap and metallic edge states at surface steps, indicating that the new structure is also a higher-order topological insulator, just like \aBiBr, in agreement with density functional theory (DFT) calculations.
\end{abstract}

\maketitle

\input{content_arXiv}

\bibliography{Literatur}

\clearpage
\pagebreak
\onecolumngrid
\input{Supplement_arXiv}

\end{document}

%% file: content_arXiv.tex
\section{Introduction}

\begin{figure*}[htb]
	\centering
	\includegraphics[]{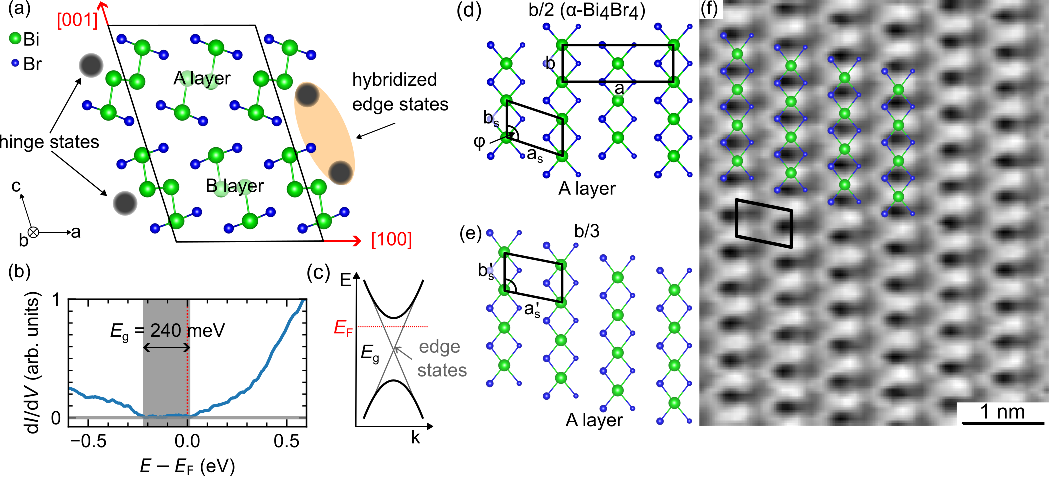} 
	\caption{(a) Atomic model of \aBiBr, with AB layer stacking. The black line shows a projection of the monoclinic bulk unit cell.  Each monolayer is a quantum spin Hall (QSH) insulator with edge states (gray circles). On the right the edge states hybridize, while on the left, they form  two hinge states. See main text for more details. (b) Scanning tunneling spectrum of the $b/3$ surface with band gap $E_\text{g}$, measured at the setpoint $V_\text{tip} = \SI{0.6}{\volt}$ and $I_\text{t} = \SI{50}{\pico \ampere}$. (c) Schematic band structure with topological gap and edge states. 
		(d) Top-view of the $\alpha$-\ch{Bi4Br4}(001)-A surface. The lower \ch{Br} atoms are indicated by smaller size.  Neighboring chains are shifted by $\mathbf{b}/2$ with respect to each other. The rectangle shows the projected monoclinic bulk unit cell (which equals the non-primitive centred surface unit cell), the parallelogram is the primitive surface unit cell. (e) The experimentally observed surface structure in which neighboring chains are shifted by $\mathbf{b}/3$ against each other. The primitive surface unit cell is shown.
		(f) STM topography recorded at $I_\text{t} = \SI{0.4}{\nano \ampere}$ and $V_\text{tip} = \SI{-0.4}{\volt}$. The atomic model from (e) is superimposed on the chains, showing good agreement. The image was upscaled using a linear interpolation.  The spectrum displayed in (b)  was smoothed using a moving average with a \SI{14}{\milli \electronvolt} window. }\label{fig:Figure1}
\end{figure*}

The quasi 1D vdW materials Bi$_4$X$_4$, where X=(Br, I), have attracted recent interest, because, amongst other aspects, their non-trivial band topologies realize exotic electronic properties  \cite{liuWeak2016, hanQuasionedimensional2022}. Monolayer \BiBr was identified both in theory and experiments to be a quantum spin Hall (QSH) insulator \cite{zhouLargeGap2014, zhouTopological2015, pengObservation2021, yangLargeGap2021},manifesting in an insulating 2D interior and gapless helically spin-polarized edge states \cite{bernevigQuantum2006, lodgeAtomically2021}, with potential applications in spintronics and quantum computing \cite{aliceaNew2012}. These properties are maintained for a monolayer of \BiBr on a bulk \aBiBr substrate, without significantly altering its QSH properties \cite{zhouTopological2015, zhouLargeGap2014}. As a result, along step edges of the monolayer, topologically protected edge states were observed, which persisted from \SI{4.2}{\kelvin } up to room temperature \cite{shumiyaEvidence2022}.

Bulk \aBiBr is a higher-order topological insulator (HOTI): 
Along lines ("hinges") where certain surface facets meet, e.g., (100) and  (001) in the case of \BiBr, topologically protected so-called hinge states can exist.  These hinge states are one-dimensional states extending along the (step) edge between two facets \cite{hossainQuantum2024, shumiyaEvidence2022, zhaoTopological2023, noguchiEvidence2021, yoonQuasiOneDimensional2020}. Bulk \aBiBr can be thought of as a stack of multiple individual monolayers, each of which would be a QSH insulator with edge states. However, due to the AB stacking of \aBiBr,  pairs of QSH edge states in adjacent layers interact in a way that is specific for the orientation of the side facet. This is illustrated in the atomic model of a \aBiBr unit cell in Fig.~\ref{fig:Figure1}(a) for a stack of two monolayers. On the right ([$100$] facet), where the two chains meet at an obtuse angle $<180^\circ$, the two edge states hybridize \cite{zhaoTopological2023}. In contrast, on the left ([$\bar 100$] facet), where the chains meet at an angle  $>180^\circ$ and consequently the two edge states are separated by protruding atoms in both chains, they do not hybridize, forming instead two separate hinge states at the top and the bottom of the stack \cite{shumiyaEvidence2022}. The name hinge state becomes clearer if one considers four- or six-monolayer stacks, where the edge states hybridize as discussed above, leaving one state at the top of the stack as a hinge between the [$100$] and [$001$] facets, and another one at the bottom and on the same side of the stack as a hinge between the [$100$] and [$00\bar1$] facets, cf.~the schematic drawings which illustrate the systematics of hinge states on terraces with (100) orientation are displayed in section~\ref{sec:Hybridisation_stacking} of the supplement. This schematics also makes it clear that the edge-state properties depend on the stacking arrangement of the layers. However, the influence of (hypothetical) shifts along the chains, i.e., the stacking of the chains in the layer plane, is not clear at present.

In this paper, we report a new  structure of the $\alpha$-\ch{Bi4Br4}(001) surface. In high-resolution STM topography images, we observe a shift of $\mathbf{b}/3$ between neighboring chains (in at least the top two layers of the sample) and propose a corresponding crystal structure. Analyzing the STM topography of terraces separated by monolayer steps, we demonstrate that the AB stacking typical for \aBiBr is preserved, despite the structural modification within the planes. Experimental spectra taken at step edges and results from our density functional theory (DFT) calculations confirm the existence of hinge states in the $b/3$-structure. We thus conclude that the $b/3$-structure also is a topological insulator.

Before turning to our results, we briefly summarize the crystal structure of \aBiBr. It crystallizes in a base-centered monoclinic structure with space group C2/m \cite{vonbendaZur1978}. The lattice constants are $a = \SI{1.3064}{\nano \metre}$, $b =  \SI{0.4338}{\nano \metre}$, $c = \SI{2.0061}{\nano \metre}$, and $\sphericalangle a c  \equiv \beta = \SI{107.42}{\degree}$ \cite{vonbendaZur1978}. Figure~\ref{fig:Figure1}(a) shows an atomic model of the monoclinic unit cell of \aBiBr projected along the $b$ axis, while Fig.~\ref{fig:Figure1}(d) displays a top view of the $\alpha$-\ch{Bi4Br4}(001) surface.  The \ch{Bi} atoms (green) form short, four-membered and $S$-shaped units in the $ac$ plane, with four \ch{Br} atoms (blue) each bonded to the top and bottom \ch{Bi} atoms. These \ch{Br} atoms connect the $S$-shaped \ch{Bi} units into infinite chains that extend in the $\mathbf{b}$ direction, cf.\ Fig.~\ref{fig:Figure1}(d). Since all inter-chain bonds are of vdW type, \aBiBr can be classified as a quasi one-dimensional vdW crystal. In Fig.~\ref{fig:Figure1}(d) both the projected bulk unit cell, which maps on a centered rectangular surface unit cell, and the primitive surface unit cell (parallelogram) of the $\alpha$-\ch{Bi4Br4}(001) surface are indicated. The primitive unit cell is spanned by lattice vectors $\mathbf{b}_\text{s}$ (with $\mathbf{b}_\text{s} = \mathbf{b}$) and $\mathbf{a}_\text{s}$ with length $a_\text{s}= \sqrt{(\nicefrac{a}{2})^2 + (\nicefrac{b}{2})^2} = \SI{0.688}{\nano \metre}$. The angle between these lattice vectors is $\varphi = \SI{107.8}{\degree}$. The inter-chain distance $d$ between two neighboring chains is given by $d = \nicefrac{a}{2} = \SI{0.653}{\nano \metre}$. As can be seen in Fig.~\ref{fig:Figure1}(a), \aBiBr exhibits AB stacking along the $c$ axis, with chains of $S$- and \reflectbox{$S$}-shaped cross section stacked on top of each other. Due to the AB stacking, the (001) surface can either expose an A or a B layer. In the projection of Fig.~\ref{fig:Figure1}(a), the layers are distinguished by the topmost \ch{Br} atom being placed either to left (A) or to the right (B) of the \ch{Bi} atom \cite{shumiyaEvidence2022, hossainQuantum2024}. The top view in Fig.~\ref{fig:Figure1}(d) shows the A surface, the lower-lying \ch{Br} atoms are indicated by a smaller diameter. The A-to-A or B-to-B step height is equal to the height of the unit cell, $h = c \sin\beta = \SI{1.91}{\nano \meter}$. Therefore, the step height from A to B (or B to A) is \SI{0.96}{\nano \meter}. Note that neighboring atomic chains on the $\alpha$-\ch{Bi4Br4}(001) surface are shifted by $\mathbf{b}/2$ with respect to each other (cf.\ Fig.\ref{fig:Figure1}(d)). This can be clearly seen in the rectangular unit cell spanned by the lattice vectors $\mathbf{a}$ and $\mathbf{b}$.

\section{Methods}

The \aBiBr single crystal, grown by the self-flux method \cite{zhaoTopological2023}, was glued to a standard sample plate using conductive epoxy. The sample was then introduced into UHV, where it was cleaved at room temperature using Kapton tape, which was peeled off along the $\mathbf{b}$ direction.

Experiments were carried out in the low-temperature four-tip scanning tunneling microscope (STM) in Oak Ridge at \SI{4.7}{\kelvin} using a commercial \ch{PtIr} tip. Scanning tunneling spectroscopy (STS) was performed with a lock-in amplifier, employing a modulation frequency $f = \SI{500}{\hertz}$ and modulation amplitudes $V_\text{mod} = \SI{10}{mV} \text{ to } \SI{100}{mV}$. Figure~\ref{fig:Figure1}(b) shows a scanning tunneling spectrum of the \aBiBr crystal used in this study (with the new $b/3$ structure), showing properties which are consistent with the literature. The bulk band gap is $E_\text{g} = \SI{240}{\milli \electronvolt}$, as estimated from multiple spectra using the method described in Ref.~\cite{lupkeLocal2022}. This value falls into the range of experimentally measured band gaps of \aBiBr reported in the literature ($b/2$ structure), either from ARPES (\SI{0.3}{\electronvolt} \cite{noguchiEvidence2021} and \SI{0.23}{\electronvolt} \cite{yangLargeGap2021}), STS (\SI{0.26}{\electronvolt} \cite{shumiyaEvidence2022, hossainQuantum2024} and $\sim \SI{0.2}{\electronvolt}$ \cite{yangLargeGap2021}), or optical measurements (\SI{0.22}{\electronvolt}) \cite{pengObservation2021}. A schematic sketch of the band structure is depicted in Fig.~\ref{fig:Figure1}(c).

Density functional theory (DFT) calculations were carried out to investigate the electronic properties and topological band character of \BiBr monolayers, both in the well-known $b/2$ and the new $b/3$ structures. 
We employed  the Vienna ab-initio simulation package (VASP) \cite{kresseEfficient1996} with the Heyd–Scuseria–Ernzerhof hybrid functional (HSE06) \cite{heydHybrid2003} to describe the exchange-correlation potential, and set the energy cutoff of the plane-wave basis to \SI{300}{\electronvolt}. To construct maximally localized Wannier functions for the $p$ orbitals of Bi and Br atoms, we used the WANNIER90 code \cite{mostofiWannier902008} and performed the calculations on a $6 \times 6 \times 3$ $k$-mesh.
Relaxing the \BiBr monolayer structure did not lead to crystal structures that are compatible with the experimental structures, which we tentatively assign to the difficulty of describing vdW interactions in DFT. 
Therefore, we fixed the unit cells for the calculations according to the literature and experimental data, respectively, without further relaxation.

A second set of DFT calculations was carried out to analyze the total energies (per unit cell) as a function of the shift between neighboring \BiBr chains in the monolayer structure.   
In these calculations, we used the experimentally determined crystal structure of monolayer \BiBr \cite{vonbendaZur1978}. Periodic boundary conditions with a \SI{2}{\nano \metre} vacuum gap between monolayers were employed. The chains were shifted rigidly against each other, without relaxation from their bulk structure in \aBiBr.  
The calculation were performed with a plane-wave basis set as implemented in Quantum ESPRESSO \cite{giannozziQUANTUM2009, giannozziAdvanced2017}. 
We used the generalized gradient approximation (GGA) with the Perdew-Burke-Ernzerhof (PBE) \cite{perdewGeneralized1996} exchange-correlation functional.  Further, we employed scalar relativistic projector augmented wave (PAW) pseudopotentials, generated using an "atomic" code \cite{dalcorsoPseudopotentials2014}
, with a non-linear core correction; semicore $d$ electrons were treated as valence electrons. The plane-wave cutoff energies for the electronic wave functions and the charge density were set to \num{60} and \SI{720}{Ry}, respectively, and the energy convergence criterion was set to \SI{e-9}{\electronvolt}. A $8\times 8 \times 2$ Monkhorst-Pack grid was chosen for $k$-point sampling. Convergence was verified for the $b/2$-, $b/3$- and $0$-shifted geometries.

\section{Experimental Results}

\subsection{Surface structure}

\begin{figure}[b]
	\centering
	\includegraphics{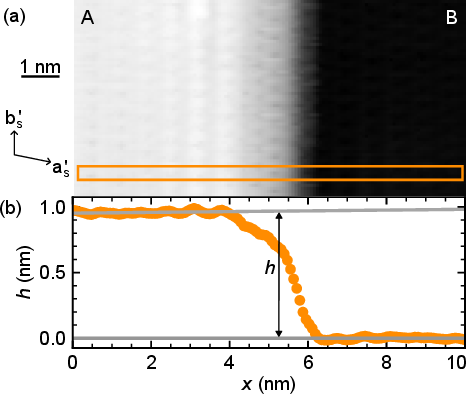}
	
	\caption{(a) STM topography of a step edge on the \ch{Bi4Br4}(001) surface, acquired at $I_\text{t} = \SI{0.1}{\nano \ampere}$ and $V_\text{tip} = \SI{-0.3}{\volt}$. The image shows a one-monolayer step edge with atomic 
		resolution on both the upper and the lower terraces. (b) Line profile indicated by the orange box in (a).
	}\label{fig:Figure2_new}
\end{figure}

\begin{figure}[tb]
	\centering
	\includegraphics[width = \linewidth]{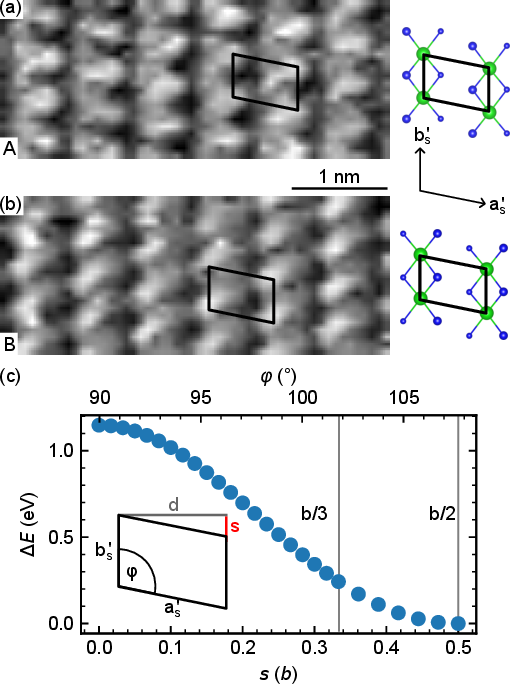}
	\caption{STM topography of the upper terrace (a) and the lower terrrace (b) in Fig.~\ref{fig:Figure2_new}. Both images were recorded at $I_\text{t} = \SI{+0.1}{\nano \ampere}$ and $V_\text{tip} = \SI{+0.4}{\volt}$.  They were upscaled using a linear interpolation. Comparing the location of the dark depression in the unit cell to the location of the lower \ch{Br} atom in the atomic models, we conclude that the upper terrace exposes an A surface, the lower terrace a B surface.  (c) Total energy per unit cell relative to the $b/2$ structure as a function of the shift $s=-d\cot\varphi$ (plotted in units of $b$) between neighboring chains, where $d$ is the inter-chain distance. $\varphi$ is the angle between the constant unit cell vector $\mathbf{b}_\text{s}$ and the changing $\mathbf{a}_\text{s}$ associated with $s$.  The inset defines $\varphi$, $d$, $s$, $a'_\text{s}$, and $b'_\text{s}$.  }
 \label{fig:Figure3_new}
\end{figure}

Figure~\ref{fig:Figure1}(f) shows an atomically resolved STM topography image of the \ch{Bi4Br4}(001) surface.
It clearly shows parallel chains, running along the $b$ axis (from top to bottom in the image). However, the mutual alignment of these chains does not correspond to the bulk-terminated $\alpha$-\ch{Bi4Br4}(001) surface:  Instead of the expected $s = b/2$ offset (cf.\ Fig.~\ref{fig:Figure1}(d)) that was already observed in the literature \cite{shumiyaEvidence2022, hossainQuantum2024, yangLargeGap2021}, the image reveals a shift of $s = b/3$ between neighboring chains. This structure is no longer spanned by the two surface lattice vectors $\mathbf{a}_\text{s}$ and $\mathbf{b}_\text{s}$, but new lattice vectors $\mathbf{a}_\text{s}'$ and $\mathbf{b}_\text{s}'$ displayed in Fig.~\ref{fig:Figure1}(e). From Fig.~\ref{fig:Figure1}(f), we extract the lengths of the lattice vectors as $a_\text{s}' = \SI{0.69(4)}{\nano \metre}$  and $b_\text{s}' = \SI{0.45 (3)}{\nano \metre}$ with an angle $\varphi = \SI{101(5)}{\degree}$ between them; the inter-chain distance is determined as $d = \SI{0.68\pm 0.04}{\nano \metre}$. These values are averages obtained from multiple STM topographs, including Fig.~\ref{fig:Figure1}(f) and \ref{fig:Figure3_new}. We note that although  the lattice vector $\mathbf{b}_\text{s}'$ appears slightly larger than that of $\alpha$-phase \ch{Bi4Br4}  ($b_\text{s} =  \SI{0.4338}{\nano \metre}$ \cite{vonbendaZur1978}), both still coincide within the experimental error of $b_\text{s}'$, i.e. $b_\text{s}' \approx b_\text{s}$.  Superimposed on the STM topography in Fig.~\ref{fig:Figure1}(f), a model of the $b/3$ structure based on rigidly shifted chains shows good agreement with the experimental image. The positions of the individual atoms on the surface were identified by comparison with a simulated STM image \cite{yangLargeGap2021} and  annotated STM topographies in Ref.~\cite{shumiyaEvidence2022}.

For several reasons, scan distortions can be excluded as the origin of the new structure. First, before the experiments on \ch{Bi4Br4}, we carefully calibrated the STM on the Au(111) surface. Second, the fact that the structure is consistent both within one image and between multiple scans of the same area rules out piezo creep and drift. Third, on terraces separated by a monolayer step the same structure rotated by 180$^{\circ}$ is observed.

On the \ch{Bi4Br4}(001) surface, we observed multiple parallel steps running in $\mathbf{b}_\text{s}'$ direction. Fig.~\ref{fig:Figure2_new}(a) shows a high resolution STM topography of such a step. From the height profile displayed in Fig.~\ref{fig:Figure2_new}(b), a step height of \SI{0.97}{\nano \metre} can be extracted. This step height compares well with experimental step heights reported in the literature \cite{yangLargeGap2021, shumiyaEvidence2022}. According to the crystal model in Fig.~\ref{fig:Figure1}(a), on $\alpha$-\ch{Bi4Br4}(001) a step height of $\SI{0.96}{\nano \metre}$ corresponds to a change from an A to a B surface or vice versa. Since both terraces in Fig.~\ref{fig:Figure2_new}(a) are atomically resolved, this image can be used to check whether AB stacking is also observed for the $b/3$ structure. Figs.~\ref{fig:Figure3_new}(a) and \ref{fig:Figure3_new}(b) display STM topographies of the same upper and lower terraces that were also imaged in  Fig.~\ref{fig:Figure2_new}. Placing parallelograms that mark the surface unit cell on both images, using the bright features as anchor points, we are indeed able to distinguish between the A and B surfaces, utilizing the location of the largest depression within the unit cell as an indicator. In Fig.~\ref{fig:Figure3_new}(a), this depression is found in the left half of the unit cell. Comparing the atomic model in Fig.~\ref{fig:Figure1}(e) and ~\ref{fig:Figure3_new}(a), the upper terrace thus exhibits the A surface. In contrast, in Fig.~\ref{fig:Figure3_new}(b) the depression is located in the right half of the unit cell, corresponding to a B surface. We therefore conclude that the $b/3$ structure exhibits the same AB stacking as \aBiBr. From this, we can also exclude that we have observed $\beta$-\BiBr, as this polymorph of \BiBr is expected to feature AA stacking \cite{liPressureinduced2019}.    

\subsection{Electronic structure}

\begin{figure*}[htb]
	\centering
	\includegraphics[]{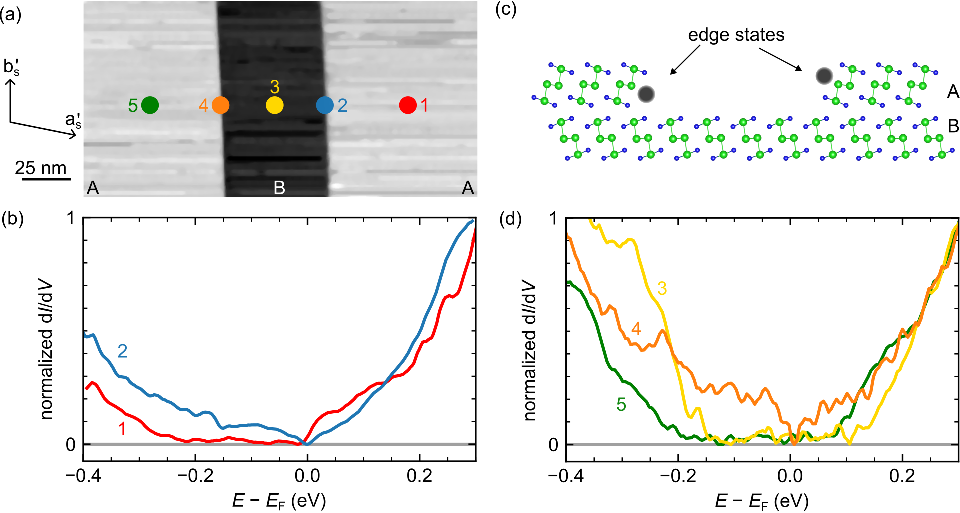}
	
	\caption{(a) STM topography of a one monolayer deep deep trench on the surface, bordered by two \BiBr monolayer steps. Scanning tunneling spectra in (b) and (d) were recorded on terraces right and left of the trench and directly at the two step edges. The filled circles indicate the approximate positions were the spectra were measured. All spectra are normalized to 1 at \SI{0.3}{\electronvolt}. Spectra at the step edges (orange and blue) show metallic edge states instead of a band gap. Note that the band gaps in the spectra from the terraces (green, yellow, and red) vary in width and alignment with the Fermi level. We explain this with sample degradation, see main text for more details. (c) An atomic model of the trench with the edge states indicated by the filled circles.   The spectra displayed in (b) and (d)  were smoothed using a moving average with a \SI{14}{\milli \electronvolt} window. }\label{fig:Figure4} 
\end{figure*}

We now turn to the electronic properties of the $b/3$ structure. Figure~\ref{fig:Figure4}(a) shows two parallel steps that together form a one monolayer deep  trench on the surface. The step to the left of this trench has already been displayed in greater detail in Fig.~\ref{fig:Figure2_new}. Figure~\ref{fig:Figure4}(b) and (d) display scanning tunneling spectra taken at five different color-coded locations in Fig.~\ref{fig:Figure4}(a). While the spectra that were recorded on the terraces both inside and outside the trench exhibit the bulk band gap with vanishing density of states (DOS), the spectra recorded at the monolayer step edges show robust metallic states in the gap,   similar to literature reports for the $b/2$ structure \cite{yangLargeGap2021, shumiyaEvidence2022, hossainQuantum2024, pengObservation2021}. We interpret this non-vanishing DOS of the $b/3$ structure within the bulk band gap as deriving from the metallic hinge states that are expected for a monolayer step on the $\alpha$-\ch{Bi4Br4}(001) surface if the monolayer is a QSH insulator (see section~\ref{sec:Hybridisation_stacking} of the supplement). We thus conclude that a \BiBr monolayer in the $b/3$ structure is a QSH insulator, just as in the $b/2$ structure.     

It is well-known that the precise location of hinge states in \aBiBr depends on the step height \cite{shumiyaEvidence2022, hossainQuantum2024, zhaoTopological2023}. Considering a trench with same step height on both sides, as in the present case, there will always be two hinge states in the trench, one at the bottom of the trench, the other at the top, no matter which facets (($100$), ($\bar{1}00$), ($10\bar{1}$), or ($\bar{1}01$)) form the walls of the trench. 
For a step height corresponding to an even number of layers, the two hinge states appear on the same side of the trench. In contrast, for an odd number of layers, they are located at opposite sides, cf. section~\ref{sec:Hybridisation_stacking} of the supplement and Refs.~\cite{shumiyaEvidence2022, hossainQuantum2024, zhaoTopological2023}. Thus, if the step height corresponds to a single layer, as in the present case, there will be hinge states on both sides of the trench (Fig.~\ref{fig:Figure4}(c)). Evidently, they are indistinguishable in tunneling spectra, unlike the opposite hinge states for trenches that are an odd number $n>1$ of layers deep. These considerations show that electronic spectra in Fig.~\ref{fig:Figure4}(b) and (d) are consistent with step height of one monolayer, although they do not allow to reach independent conclusions regarding the stacking sequence.

We note that the spectra labelled 1, 3, and 5 that were recorded on the three terraces in Fig.~\ref{fig:Figure4} show slight differences in width and position of the bulk band gap. We tentatively attribute this effect to a degeneration of the surface due to accumulation of adsorbates over the duration of the experiments (5 days), which lead to a hole doping and ultimately metallic spectra.

\begin{figure*}
    \centering
	\includegraphics[width = \linewidth]{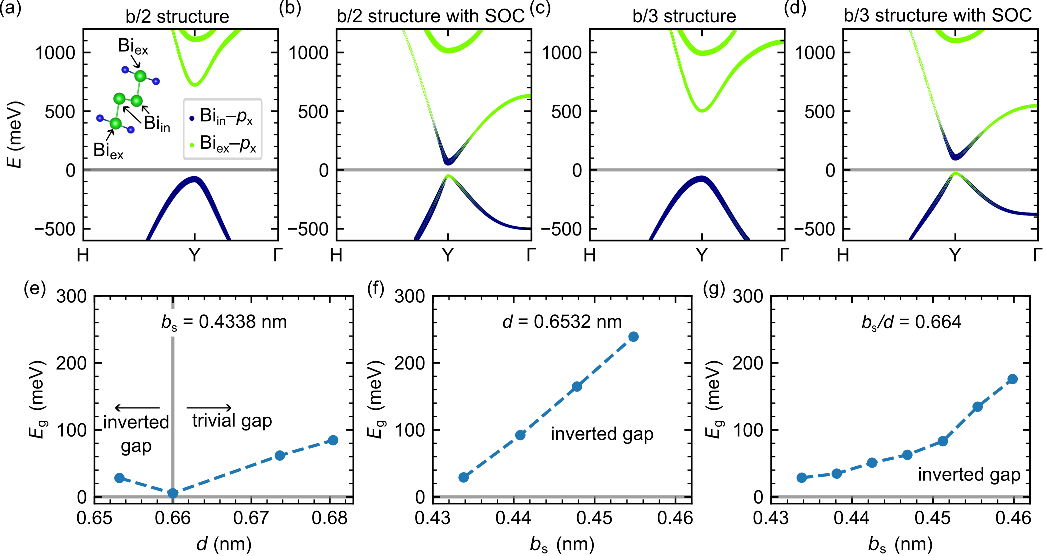}
	\caption{(a) and (b) show the \ch{Bi}$_\text{in}$-$p_x$ and  \ch{Bi}$_\text{ex}$-$p_x$ projected orbital character of the conduction and the valence bands around the Y point without (a) and with (b) SOC for the $b/2$ structure, see \cite{zhouLargeGap2014}. (c) and (d) show the same projected orbital character of the bands around the Y point for the $b/3$ structure. (c) and (d) were calculated using the lattice constants indicated in the main text.  For both structures, SOC exchanges the two orbitals leading to an exchange of parity, which makes both the $b/2$ (b) and the $b/3$ (d) structure a QSH insulator. 
 (e) - (g): evolution of the inverted band gap $E_\text{g}$ as a function of the inter-chain distance $d$ and the lattice constant $b_\text{g}$. The dashed lines are a guide for the eyes. (e) Dependence of $E_\text{g}$ on the inter-chain distance $d$, when fixing $b = \SI{0.4338}{\nano \metre}$. The inverted band gap decreases, until it closes at $d = \SI{0.66}{\nano \metre}$ and then transitions to a trivial gap. (f) Fixing $d = \SI{0.6532}{\nano \metre}$ and increasing the lattice vector $b_\text{s}$ enhances the the inverted band gap. (g) The same applies if the ratio $b_\text{s}/d = \num{0.664}$ is kept constant.
 }
 \label{fig:Fig_DFT}
\end{figure*}

We now analyse with DFT whether the new $b/3$ structure is expected to be a QSH insulator. It is well-known that DFT predicts the \BiBr monolayer in the $b/2$ structure to be a  QSH insulator \cite{zhouLargeGap2014, zhouTopological2015}.
For the $b/2$ structure, the QSH property arises from the inverted band gap at the Y point that is caused by the spin-orbit coupling (SOC). The latter exchanges the two \ch{Bi}-$6p$ orbitals from which the lowest conduction band and the highest valence band derive. Crucially, this also leads to the exchange of parity between the conduction and valence bands at the Y  point (Fig.~\ref{fig:Fig_DFT}(a) and \ref{fig:Fig_DFT}(b)), which makes the $b/2$ structure a QSH insulator.

Since the experimentally determined inter-chain distance $d'$ and the surface lattice vector $\mathbf{b}_\text{s}'$ of the $b/3$ structure are somewhat larger than the literature values for \aBiBr, and since it is evident that these structural parameters must have a strong influence on the band structure, we investigated the evolution of the band gap with $d'$ and $b_\text{s}'$. Fig.~\ref{fig:Fig_DFT}(e) displays the dependence of the nature and size of the  band gap of the $b/3$ structure on $d'$ when $b_\text{s}'$ is fixed to the literature value of the $b/2$ structure, $b_\text{s}=\SI{0.4338}{\nano \metre}$. We find that the $b/3$ structure becomes a trivial insulator for $d' > \SI{0.66}{\nano \metre}$. Since this threshold still lies within the error bars of the experimentally determined value $d' = \SI{0.68(4)}{\nano \metre}$, the DFT result is compatible with the $b/3$ structure indeed being a QSH insulator. We note that the inter-chain distance of the $b/2$ structure ($d=\SI{0.653}{\nano \metre}$) is well below the threshold of $\SI{0.66}{\nano \metre}$.  
However, the experimental $b_\text{s}'$ is larger than the literature value $b_\text{s}$ for the $b/2$ structure. Therefore, we also investigate the dependence of the topological properties on $b_\text{s}'$. To this end, we fix $d'$ at $d=\SI{0.6532}{\nano \metre}$, i.e., the value for the $b/2$ structure, and increase $b_\text{s}'$, starting from its literature value for the $b/2$ structure ($b_\text{s} = \SI{0.4338}{\nano \metre}$). The result is displayed in Fig.~\ref{fig:Fig_DFT}(f). It shows that stretching $b_\text{s}'$ has the opposite effect as stretching $d'$, i.e., a larger $b_\text{s}'$ stabilizes the QSH property by increasing the inverted gap, and that this effect is significant. It should stabilize the non-trivial topological properties of the $b/3$ structure with its larger $b_\text{s}'$ even further. 

The opposite scaling behavior of topological properties  with  $d'$ and $b_\text{s}$ raises the question which of these two effects is stronger. Therefore, we investigated the nature and size of the gap as   both parameters, $d'$ and $b_\text{s}'$, are increased simultaneously, such that their ratio is constant (clearly, this also requires a scaling of $a_\text{s}'$). Since ---interestingly---this ratio is nearly identical for the $b/2$ and $b/3$ structures ($ b_\text{s}'/d'=0.662$ and $ b_\text{s}/d=0.664$, we fixed the ratio to  $0.664$. The result is plotted in Fig.~\ref{fig:Fig_DFT}(g). In the complete range the gap is inverted and the $b/3$ structure is a QSH insulator, the gap of which opens further with increasing $b_\text{s}'$. Hence, the larger experimental value of $b_\text{s}'>b_\text{s}$ stabilizes the QSH insulator property, even if the mutual displacement along $\mathbf{b}_\text{s}$ of the chains in the $b/3$ structure for steric reasons (stronger repulsion between Br atoms) should lead to an increased (and destabilizing) $d'>d$. 

From the structures represented in Fig.~\ref{fig:Fig_DFT}(g), we find the closest agreement with the experimental parameter set ($a_\text{s}'=\SI{0.69(4)}{\nano \metre}$, $b_\text{s}'=\SI{0.45 (3)}{\nano \metre}$, $d=\SI{0.68\pm 0.04}{\nano \metre}$, and $\varphi = \SI{101(5)}{\degree}$) for the fifth data point from the left, with   calculated parameters $a_\text{s,c}' = \SI{0.6959}{\nano \metre}$, $b_\text{s,c}' = \SI{0.4512}{\nano \metre}$, $d_\text{c}' = \SI{0.6795}{\nano \metre}$, $\varphi_\text{c}' = \SI{102.48}{\degree}$. For this model of the $b/3$ structure, the energy gap is predicted at $E_\text{g}' = \SI{80}{\milli \electronvolt}$. This is smaller than the gap DFT-predicted gap of the $b/2$ structure ($E_\text{g} = \SI{100}{\milli \electronvolt}$). However, it should be noted in this context that DFT generally tends to underestimate band gaps. For the $b/2$ structure, experimentally determined gaps are in the range \SIrange{200}{300}{\milli \electronvolt}; for STS measurements on the surface of a bulk \aBiBr crystal in the $b/2$ structure $E_\text{g} =\SI{260}{\milli \electronvolt} $ has been reported \cite{shumiyaEvidence2022, hossainQuantum2024}. In our STM experiments, we observe a slightly smaller gap of $E_\text{g}' =\SI{240}{\milli \electronvolt}$. This is line with a slightly smaller DFT-predicted gap of the $b/3$ structure when compared with the $b/2$ structure, suggesting that the former is somewhat closer to the transition to trivial topology. Figure~\ref{fig:Fig_DFT}(c) and \ref{fig:Fig_DFT}(d) show Bi$_\text{in}$-$p_x$ and Bi$_\text{ex}$-$p_x$ orbital projected character of bands without and with SOC, respectively, in analogy to Figs.~\ref{fig:Fig_DFT}(a) and \ref{fig:Fig_DFT}(b) for the $b/2$ structure. Also for the $b/3$ structure, the SOC leads to a change in parity at the Y point. We can therefore conclude that it, too, is a  $\mathbb{Z}_2 = 1$ QSH insulator.

Finally, we remark that, alternatively, one can take the viewpoint that a change in $b_\text{s}$ from the $b/2$ to the $b/3$ is unlikely, since this lattice constant is determined by covalent bonds. In that case, we define $b_\text{s}'\equiv b_\text{s}=\SI{0.4338}{\nano \metre}$ and scale the experimentally measured $a_\text{s}'$  and $d'$ accordingly (assuming an isotropic distortion of the experimental image in Fig.~\ref{fig:Figure1}(f)), obtaining $\tilde{a}_\text{s}'=\SI{0.67(4)}{\nano \metre}$ and $\tilde{d}'=0.66(4)$. The latter value is close to the calculated threshold for a non-trivial topology ($d'=0.66$), but given the experimental error, the result is once more compatible with the $b/3$ structure being a QSH insulator, although again somewhat weaker than the $b/2$ structure.

\section{Discussion}
We now turn to a discussion of the origin of the observed $b/3$ structure, which has not been reported in the literature so far.
Since  STM only allows access to the topmost layer, we cannot determine if the $b/3$ crystal structure is present deeper in the sample or only at its surface.
However, since we observe the $b/3$ lattice to be present across step edges (i.e., on adjacent surface terraces),  we conclude that either at least the top two monolayers of our sample are in the $b/3$ structure,  or that the topmost layer systematically exhibits the $b/3$ structure as a kind of surface reconstruction. The latter possibility is rendered unlikely by the fact that other authors have observed the $b/2$ structure at the surface. We thus conclude that in the present case at least the topmost two layers show the new structure, because we consider a coincidental change below the step edge to be unlikely.

The possibility of a surface reconstruction of the uppermost layer is also essentially ruled out by a DFT calculation. Fig.~\ref{fig:Figure3_new}(c) displays the total energy $\Delta E_{\rm tot}$ per monolayer unit cell as function of the shift $s$ in units of $b$,  referenced to $E_{\rm tot}$ of the $b/2$ structure.
Also shown is the dependence on the angle $\varphi$ between the two primitive unit cell vectors $\mathbf{a}_\text{s}$ and $\mathbf{b}_\text{s}$.
Evidently, the $b/2$ structure ($s=\nicefrac{b}{2}$ and $\varphi =\SI{107.8}{\degree}$) represents an energetic minimum, with the total energy increasing monotonically towards the $b/3$ structure with $s=\nicefrac{b}{3}$ and $\varphi= \SI{102}{\degree}$.
The energetically least favorable configuration is $s=0$ ($\varphi = \SI{90}{\degree}$).
Overall, the behavior in Fig.~\ref{fig:Figure3_new}(c) is to be expected by steric hinderance between atoms in neighboring chains. While these results suggest that the $b/3$ structure is energetically not favorable and does not even correspond to a local energy minimum, it should be kept in mind that the DFT calculations were carried out for a single \BiBr monolayer of the bulk structure and moreover with rigid chains (no relaxation of the internal chain structure or the experimental inter-chain distance of $\SI{0.677}{\nano \metre}$). In principle, distortions within the chains could stabilize the $b/3$ structure. However, we consider it unlikely that a total energy increase of 0.25\,eV (Fig.~\ref{fig:Figure3_new}(c)) per unit cell could be offset by such intra-chain distortions. We thus finally reject the idea of a surface reconstruction that occurs in order minimize the surface energy.

This leaves us with the scenario of a strained crystal. Because of the quasi 1D nature of \aBiBr, it is conceivable that externally applied shear stress will results in chains gliding past each other. For example, it is well-known that external stress applied parallel to the planes can result in the straining of 2D vdW materials, which leads to glide shifts of the planes and significantly modifies the electronic structure \cite{sugawaraSelective2018}. 
Since in quasi 1D vdW materials the in-plane shear modulus is much smaller than in 2D compounds, because of the additional vdW gaps in the 2D plane, glide shifts of the chains are easily conceivable. 
We note that the function $\Delta E_{\rm tot} (s)$, by symmetry, is periodic $\Delta E_{\rm tot} (s)=\Delta E_{\rm tot} (s-b)$, with maximum shear strain $\gamma=(\nicefrac{b}{2}-s)/d=33\%$ at $\varphi=90^{\circ}$.
Since in quasi 1D vdW materials the in-plane shear modulus is much smaller than in 2D compounds, because of the additional vdW gaps in the 2D plane, glide shifts of the chains are easily conceivable. 
We note that the function $\Delta E_{\rm tot} (s)$, by symmetry, is periodic $\Delta E_{\rm tot} (s)=\Delta E_{\rm tot} (s-b)$, with maximum shear strain $\gamma=(\nicefrac{b}{2}-s)/d=33\%$ at $\varphi=90^{\circ}$.
This means that a strain that would lead to $s>b$ (even $s\gg b$) causes a plastic deformation with  neighboring chains shifting against each other by one or more lattice constants $b$, until the shift $s$ between neighboring rows is $0< s <\nicefrac{b}{2}$. Evidently, from our experimental data, we can only detect this residual strain.
In this scenario of a strained crystal, there is \textit{a priori} no reason why the residual strain should lock at any particular value of $s$, e.g., $s=\nicefrac{1}{3}$ (corresponding to the $b/3$ structure and $\gamma\approx7.5\%$) as in the present case. However, it should be noted that during the release of macroscopically enforced strain energy, small relaxations of the internal chain structure and thus the inter-chain interaction energy can make certain values of $s$ metastable, even if the energy gain of such relaxations is too small to globally stabilize a reconstruction of the \textit{unstrained} crystal surface at this $s$.

The strain-based explanation of the $b/3$ structure raises the question of the origin of the strain. Here, three possibilities are conceivable. First, stress could have been applied and the strain induced when handling the bulk crystal, e.g., when gluing it on the STM sample holder. In this case, we would expect the $b/3$ structure to extend throughout most of the sample.  Second, the strain could have been introduced during the cleaving process of the sample in UHV. Note that we cleaved the sample in the $\mathbf{b}$ direction, such that a small misalignment of the pulling direction from $\mathbf{b}$ may well have caused the observed strain. In both of these scenarios, the strain field would be expected to extend through the sample on macroscopic length scales. This is different in the third scenario, according to which the stress is applied on a mesoscopic or even microscopic scale by inhomogeneities in the bulk crystal, such as mosaic spread or domain boundaries. However, in our experiments we observed the $b/3$ structure throughout the whole investigated area.

\section{Conclusion}

We report a new monolayer structure of \BiBr on \aBiBr in which the shift between quasi 1D chains is $b/3$ instead of $b/2$, where $b$ is the length of the unit cell in the direction of the chains. We can exclude the possibility that this structure originates from a surface reconstruction. Instead, it is a strained structure. The strain leads to the gliding of the quasi 1D chains against each other, it does not change the AB layer stacking of the \aBiBr crystal. Experimentally, we find strong evidence that the new structure is also  a QSH insulator, because we detect edge states at monolayer steps, as expected for the higher order topological insulator \aBiBr.  A DFT calculation of the new structure supports this, finding an inverted energy gap at the Y point, opened by spin-orbit coupling which exchanges the parity between the conduction and valence bands at Y, in full analogy of the conventional $b/2$ structure. The qualitatively identical electronic structure of the $b/2$ and $b/3$ structures is remarkable, as it is known that the stacking order of 2D materials can have a major influence on their electronic and topological properties \cite{sugawaraSelective2018, lupkeQuantum2022}. Here, a massive change in the stacking of the 1D chains in the 2D \BiBr plane does not change the topology, which thus turns out to be robust, although both the experimentally determined  band gap energy and the DFT calculations with structural parameters derived from the STM experiment consistently indicate that the $b/3$ structure is closer to the transition to trivial topology than the $b/2$ structure is. Finally, our observations differ significantly from the structural and electronic properties of \aBiBr under hydrostatic pressure, which has been reported to be metallic/superconducting \cite{liPressureinduced2019}.
The corresponding triclinic unit cell observed at high hydrostatic pressure \cite{liPressureinduced2019} is also clearly distinct from our observation of AB stacking.

\section{Data Availability}

Raw data is available from the corresponding authors upon reasonable request. \\

\section{Acknowledgement}

The STM experiment was conducted at the Center for Nanophase Materials Sciences, which is a DOE Office of Science User Facility.
F.L. and F.S.T. acknowledge funding from the Bavarian Ministry of Economic Affairs, Regional Development and Energy within Bavaria’s High-Tech Agenda Project ''Bausteine f\"ur das Quantencomputing auf Basis topologischer Materialien mit experimentellen und theoretischen Ans\"atzen'' and Germany’s Excellence Strategy - Cluster of Excellence Matter and Light for Quantum Computing (ML4Q).
F.L. acknowledges funding by the Deutsche Forschungsgemeinschaft (DFG, German Research Foundation) within the Priority Programme SPP 2244 (Project No.\ 443416235), as well as the Emmy Noether Programme (Project No. 511561801).
J.J.Z acknowledges support from the National Key R\&D Program of China (Grant No. 2022YFA1403400) and the National Natural Science Foundation of China (Grant No. 12104039). 
Z.W. was supported by the National Key R\&D Program of China (No. 2020YFA0308800, and No. 2022YFA1403400), the National Natural Science Foundation of China (No. 92065109), the the Beijing National Laboratory for Condensed Matter Physics (Grant No. 2023BNLCMPKF007), and the Beijing Natural Science Foundation (Grant No. Z210006). Z.W. thanks the Analysis \& Testing Center at BIT for assistance in facility support.

%% file: Supplement_arXiv.tex

\makeatletter
\renewcommand \thesection{S\@arabic\c@section}
\renewcommand\thetable{S\@arabic\c@table}
\renewcommand \thefigure{S\@arabic\c@figure}
\makeatother
\setcounter{figure}{0}
\setcounter{section}{0}

\begin{center}
	\Large	\textbf{SUPPLEMENTAL MATERIALS: Shear-resistant topology in quasi one-dimensional van der Waals material \ch{Bi4Br4} }
\end{center}

\section{Hinge states in the HOTI \aBiBr} \label{sec:Hybridisation_stacking}

\begin{figure*}[hbp]
\centering\includegraphics{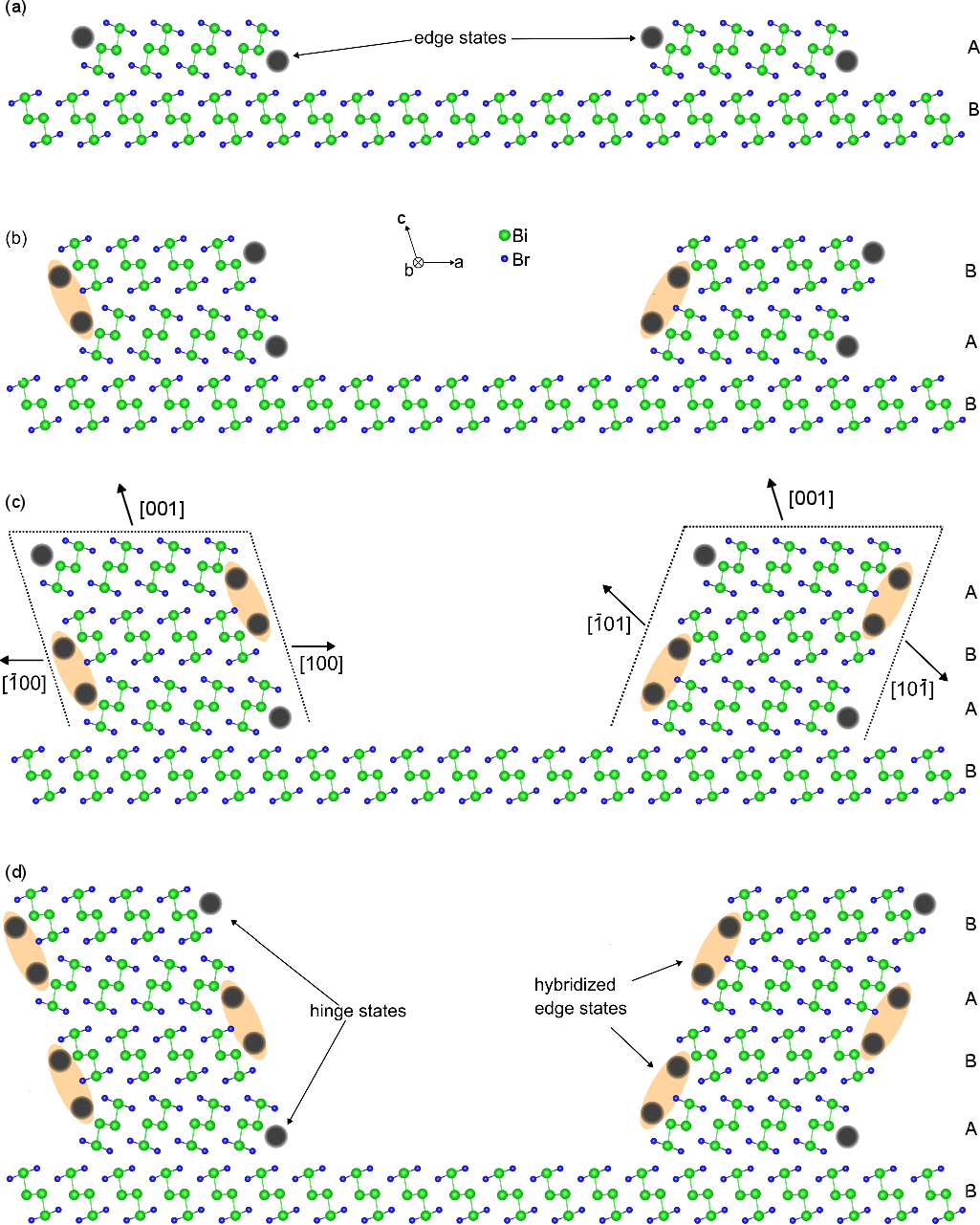}
    \caption{Formation of hinge states (filled grey circles) in AB-stacked \BiBr monolayers of varying thickness (one to four layers) on \aBiBr. The sides  of the left stack are \{$100$\}  facets, while the sides  of the right stack are \{$10\bar{1}$\} facets.  (a) Single-monolayer stack. (b) Two-monolayer stack. (c) Three-monolayer stack. (d) Four-monolayer stack. In panel (c), the side and top facets of the two stacks are indicated by dashed lines. The figure illustrates how the hybridization (orange shading) of the edge states of the individual QSH layers leads to the formation of hinge states (see \cite{shumiyaEvidence2022, hossainQuantum2024, zhaoTopological2023}).
    }
    \label{fig:Supp_HingeStateStacking}
\end{figure*}

Fig.~\ref{fig:Supp_HingeStateStacking} illustrates the formation of hinge states in AB-stacked \BiBr monolayers \cite{shumiyaEvidence2022, hossainQuantum2024, zhaoTopological2023} of varying thickness (one to four layers) on \aBiBr.
In each panel of Fig.~\ref{fig:Supp_HingeStateStacking}, two stacks with different low-index side facets are shown. For the left stack, ($100$) and ($\bar{1}00$) side facets were chosen. In contrast, the right stack has ($10\bar{1}$) and ($\bar{1}01$) side facets. In Fig.\ref{fig:Supp_HingeStateStacking}(c) these side facets are indicated by dashed lines, and the corresponding crystallographic directions, based on the monoclinic cell, are given. The figure shows that on all four side facets, adjacent edge states can hybridize, if they are not separated by protruding atoms in both chains. This leads to the emergence of pairs of hinge states at the top and bottom of each stack.  If the stack has an even (odd) number of layers, the two hinge states appear on equal (opposite) side facets. 
Due to the patterns formed by the hybridizing edge states, the two surfaces of \{$100$\} class are no longer equivalent. The same is true of the \{$10\bar{1}$\} surfaces.  

Fig.~\ref{fig:Supp_HingeStateStacking} also allows to determine the location of hinge states on either side of trenches aligned with the $[010]$ direction, for the four possible combinations of low-index side facets: (1) left $(100)$,  right $(\bar{1}01)$, (2) left $(10\bar{1})$,  right $(\bar{1}00)$, (3) left $(100)$,  right $(\bar{1}00)$, (4) left $(10\bar{1})$,  right$ (\bar{1}01)$. In all cases, the arrangenment of hinge states is qualitatively the same.

\clearpage